\documentclass[runningheads,a4paper]{llncs}

\usepackage{amssymb}
\usepackage{graphicx}
\usepackage{color}
\usepackage[caption=false]{subfig}
\usepackage{url}
\usepackage{multirow}
\usepackage{amsmath,systeme}
\usepackage{caption}

\newcommand{\keywords}[1]{\par\addvspace\baselineskip
\noindent\keywordname\enspace\ignorespaces#1}

\begin{document}

\mainmatter  

\title{Indoor Localization Techniques Within a Home Monitoring Platform}

\titlerunning{Indoor Localization Techniques Within a Home Monitoring Platform}

%
%
\author{Iuliana Marin \and Maria-Iuliana Bocicor \and Arthur-Jozsef Molnar}

%

\institute{S.C. Info World S.R.L.,\\
Bucharest, Romania\\
\email{\{iuliana.marin,iuliana.bocicor,arthur.molnar\}@infoworld.ro}
\url{https://www.infoworld.ro/en/}
}

%
%

\toctitle{Indoor localization techniques within a home monitoring platform}
\tocauthor{Marin, Bocicor, Molnar}
\maketitle

\begin{abstract}
This paper details a number of indoor localization techniques developed for real-time monitoring of older adults. These were developed within the framework of the i-Light research project that was funded by the European Union. The project targeted the development and initial evaluation of a configurable and cost-effective cyber-physical system for monitoring the safety of older adults who are living in their own homes. Localization hardware consists of a number of custom-developed devices that replace existing luminaires. In addition to lighting capabilities, they measure the strength of a Bluetooth Low Energy signal emitted by a wearable device on the user. Readings are recorded in real time and sent to a software server for analysis. We present a comparative evaluation of the accuracy achieved by several server-side algorithms, including Kalman filtering, a look-back heuristic as well as a neural network-based approach. It is known that approaches based on measuring signal strength are sensitive to the placement of walls, construction materials used, the presence of doors as well as existing furniture. As such, we evaluate the proposed approaches in two separate locations having distinct building characteristics. We show that the proposed techniques improve the accuracy of localization. As the final step, we evaluate our results against comparable existing approaches.

\keywords{indoor localization, received signal strength, trilateration, kalman filter, neural network}
\end{abstract}

\section{Introduction}
The developed world is on the cusp of long-term societal and demographic change heralded by population ageing. The World Health Organisation estimates the number of adults over 60 to double worldwide by 2050 \cite{who15}. However, the report finds that advances in healthcare and medicine do not necessarily translate into improved quality of life for older adults. This situation is expected to increase the toll on healthcare system and local government expenditures for care programs. Meijer et al. \cite{meijer2013} find that real annual health expenditure increases at 4\% per year, with an important part of these funds to be geared towards the older adult population in the future.

However, advancement in the form of diminutive computing platforms, advanced wireless networks and a push towards wearable devices provide an opportunity for technological solutions that supplement healthcare-based measures in older adult care. Special importance is given to systems that enable adults to continue living in their own homes, maintaining good social relations. The European Union identified the increased importance of caring for older adults and created the Ambient and Assisted Living Programme \cite{aal16} to help public and private organizations develop and bring solutions to market.

Our presented work is part of the i-Light research project funded by the European Union. Its main objective concerns the development of an extensible and cost-effective cyber-physical platform for home monitoring and assisted living \cite{marin2018}. The main target group are older adults living in their own homes. The hardware side of the platform is represented by a number of intelligent luminaire devices that were developed as part of the project. In addition to lighting, they provide sensing, localization and communication systems using Bluetooth Low Energy and WiFi. The developed luminaires  replace existing light bulbs and can communicate between them as well as with a remotely-deployed software server. This approach addresses adoption barriers by reducing deployment costs, simplifying installation and being inconspicuous within the home. Furthermore, a single cloud-based server deployment can service many household deployments, improving cost-effectiveness for additional installations.

The most important innovative aspects as well as platform architecture and main components were already detailed in previous work \cite{iccp2017,marin2018,marin19}. As such, this paper is focused on presenting and evaluating the implemented localization algorithms. The problem of person localization can be partitioned into outdoor and indoor localization. The former is covered by existing and well-known technologies. The first to be made available to the public was the United States' GPS system. Recognising the strategic importance of localization, additional such systems were implemented by the European Union (Galileo), Russia (GLONASS), China (BeiDou-2) and Japan (QZSS). Currently, these latter systems are focused to provide good accuracy within the strategic region of its implementing country. In order to work, all these systems require the device to have direct line of sight toward several satellites in the constellation. Indoors, these systems range between inaccurate at best and completely inoperative. As such, achieving accurate indoor localization required the development of new technologies and algorithms. These include communication via visible light \cite{haigh14}, acoustic background fingerprinting \cite{tarzia11} and user movements \cite{tarrio11}. Given the project requirements detailed in \cite{draghici17}, the selected approach was trilateration of the signal strength received from a Bluetooth Low Energy device by at least three intelligent luminaires. This has the advantage of not requiring additional wiring or unsightly devices. System accuracy was evaluated within two locations. The first one was a small home with thick concrete and brick walls \cite{marin19}, and the second one an office building having larger rooms, but fewer and thinner walls. In both cases, the minimum of three devices were deployed.

\section{Related Work}
Previously described trends have lead to an increase in the number of smart home devices in use, both in homes as well as the workplace and public places. Indoor localization is one of their strong-suits, as it allows monitoring and analysis of visitor behaviour, building customer profiles and understanding patterns of human interaction. 

As detailed in the previous section, GPS is the default localization technology used outdoors. Given that satellite signal is easily blocked by buildings \cite{ozsoy2013}, several alternative technologies were considered and evaluated for indoor localization. Among them are acoustic and optical signals, radio-frequency identification (RFID), as well as using WiFi and Bluetooth signals for triangulation or trilateration \cite{ta2017,lymberopoulos2015,xiao2016}. When electromagnetic waves are used, the distance between the target of monitoring and a number of beacons is calculated based on the Received Signal Strength Index (RSSI) \cite{huang2019}. In most cases, the beacons are fixed in known locations, and they record the strength of an electromagnetic signal received from the tracked emitter. Due to their ubiquity, a WiFi or Bluetooth signal is usually employed. The received power decreases as the distance the signal must travel increases, using the inverse square law \cite{luo2011,nobles2011,sadowski2018,maduskar2017}. This allows calculating the distance between the signal emitter and the beacons receiving the signal. Using at least three fixed beacons and applying triangulation or trilateration allows calculating the emitter's physical location. 

However, raw RSSI readings can be misleading due to multipath fading, the degree of variance being higher indoors \cite{pu2011}. This is exacerbated by the presence of walls and large items of furniture that affect signal strength. Also, for many use cases, it is desirable to carry out localization without equipping the monitored person with a wearable device, leading to continued research interest in the domain.

One of the proposed solutions to address inherent variance in readings is to apply a filter on the raw RSSI values. The Simultaneous Localization and Configuration algorithm \cite{bulten2016} is based on the fast simultaneous localization and mapping problem for RSSI-based localization and can run on a mobile device. Another filter-based approach is the Kalman filter \cite{Welch1995}, which is generally used to smooth noisy data. The Kalman filter is amenable for smoothing raw RSSI readings and has low computational overhead. It is a recursive filter that evaluates a system's state starting from a series of noisy measurements. Each new measurement is subjected to a weighted mean calculated based on the covariance for reducing uncertainty. After a number of readings, the mean and covariance are recalculated. Kalman filtering can increase the accuracy of indoor localization \cite{robesaat2017} by lowering accumulated errors \cite{sung2016}. Furthermore, \cite{sung2016} showed that filtering lowered energy consumption and enhanced the stability of the readings.

The beacons used in our paper are intelligent luminaires \cite{marin2018} that have WiFi and Bluetooth communication capabilities and which can replace existing light bulbs. They were designed to ensure that luminaire-related functionality does not impede indoor localization by influencing signal quality or directionality. The luminaires can be mounted into existing wall sockets. In our experiments, they replaced ceiling-mounted light bulbs, which provided good signal directionality and the advantage that the system looked inconspicuous. Our experiments were carried out using a smartphone emitting Bluetooth signal recorded and timestamped by the luminaires. Other Bluetooth compatible devices, such as smartwatches and emergency buttons can also be used. Our solution's primary drivers were cost-effectiveness and unobtrusiveness. The proposed setup leads to low costs associated with deployment and maintenance, and does not look inconspicuous when installed in a residential building or office.

\section{Platform Overview}
\begin{figure}[h]
    \centering
    \includegraphics[width=0.75\linewidth]{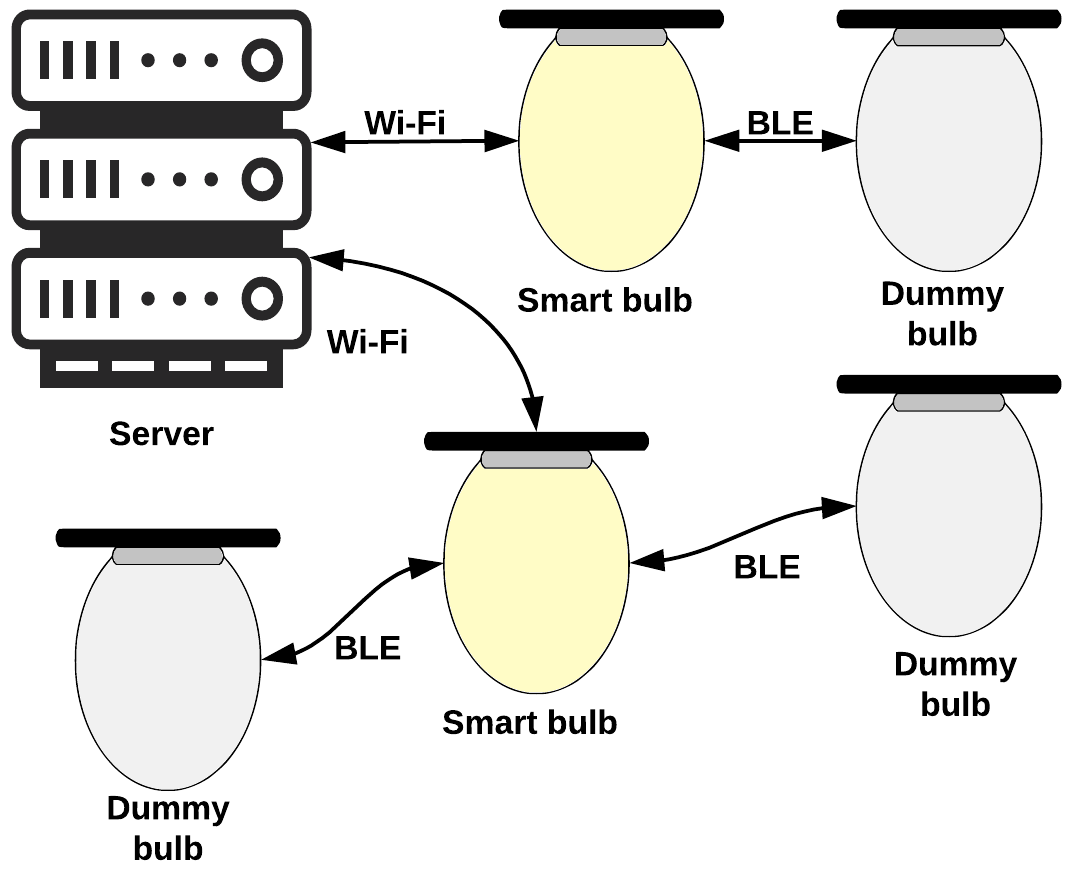}
    \caption{High-level system architecture \cite{marin19}}
    \label{fig:hardware_architecture}
\end{figure}

The i-Light platform was designed to be a home monitoring system for older adults and relies on energetically efficient, intelligent luminaires, equipped with an integrated electronic sensor system, indoor localization and communications that allow continuous, ubiquitous and inexpensive home monitoring. The high-level system architecture is presented in Figure \ref{fig:hardware_architecture} \cite{marin19}. The hardware subsystem is represented by a wireless network of luminaires, composed of two types of nodes designated as \textit{smart} and \textit{dummy}. The software subsystem integrates several modules, distributed on different devices: smart and dummy luminaires, software server and client devices. Both aforementioned subsystems will be expounded in the following subsections.

\subsection{Wireless network of luminaires}
The luminaire network is deployed within personal homes to monitor the indoor environment and track the older adult's location. Together, the installed luminaires form a network that completely covers the indoor environment. All location and ambient monitoring data is collected and sent in real time to the software server, which computes the person's indoor location and analyses environmental information continually.

\subsubsection{Smart Luminaires.}
Within the wireless luminaire network smart bulbs represent the resourceful constituents. Using a Raspberry Pi3 board \cite{halfacree12} at their core, this type of luminaire offer features such as lighting, ambient sensing and direct communication to the server and with the dummy bulbs. Their design is modular, as this has proved to be the best option to easily isolate manufacturing problems and make changes without rebuilding the entire device. With regard to lighting, the luminaires provide LED intensity management. They also include a sensor module for environment monitoring, which measures temperature, ambient light, humidity, CO\textsubscript{2}, dust and volatile organic compound gases. All values measured are sent directly to the software server via WiFi or Ethernet, where they are subsequently analysed. Smart bulbs can also identify Bluetooth-equipped devices and record RSSI values. Furthermore, they can collect RSSI values registered by several dummy luminaires and forward these to the server, where the localization algorithms are run. At system configuration, each of the dummy bulbs must be associated to a smart bulb. This design enables smart luminaires to directly communicate via Bluetooth with their associated dummy bulbs and collect localization data.

\subsubsection{Dummy Luminaires.}
Dummy luminaires are simpler, smaller and lower-cost when compared with smart luminaires. Their limited functionality allows using them for lighting and localization. They rely on a Bluegiga BLE112 Bluetooth Low Energy module \cite{ble117}, which enables them to establish a connection to the smart bulb they are associated with. Dummy bulbs have lighting capabilities, with the light intensity being controlled via the smart luminaire they are connected to. Dummy bulbs also carry out environment scans in order to detect other Bluetooth devices and acquire RSSI values. Localization is carried out on the server according to collected RSSIs. Each dummy luminaire must be associated to a smart one, as the dummies are not WiFi equipped and cannot directly send data to the server. Dummy luminaires stay connected to their associated smart bulb using Bluetooth and transfer the RSSIs in real time. The main advantage of using dummy bulbs is that the system's overall cost is reduced, while sufficient coverage is possible for accurate localization.

\subsubsection{Communication Protocol.}
Communication between smart and dummy light bulbs is achieved via Bluetooth Low Energy. For dummy bulbs, this is provided by the Bluegiga BLE112 module, while the smart bulb employs its onboard Raspberry Pi3. Communication between smart luminaires and the server is accomplished through web services with information transmitted in JSON format.

An important concern in two-way Bluetooth Low Energy connectivity is the master and slave connection roles. Slave devices advertise themselves and wait for connections. Master devices scan the environment and initiate connections to slaves. Smart luminaires act as masters, as after they perform a scan for both Bluetooth devices and dummy luminaires, they initiate connections to both: to Bluetooth devices in order to acquire RSSI values for localization and to dummy bulbs in order to receive the RSSIs collected by them. In order to send data from smart to dummy luminaires, dummy bulbs act as slaves, emitting advertisement packets allowing them to be identified by the smart bulbs. However, in order to detect additional devices in the environment, such as the actual device used for indoor positioning, they must change their role and act as masters. Consequently, dummy luminaires periodically change roles. All data acquired by both luminaire types are sent to the system server and stored in the database, where  ambient and RSSI data are processed, localization algorithms executed and real-time alerts are generated, if necessary. 

Communication between smart luminaires and the server is accomplished using the MQTT messaging protocol\footnote{http://mqtt.org/}. Bidirectional communication using JSONs is achieved via the publish/subscribe model. This ensures loose coupling between the monitoring system's components. The protocol is characterised by low bandwidth and energy consumption, reliability, and is dedicated to Internet of Things applications. Every smart bulb publishes JSONs using a subject. The subject represents routing information. The server subscribes to several subjects and analyses the received JSONs. The security of sent and received JSON messages is ensured by encryption.

\subsection{Software Server Components}
The i-Light software system includes a server and a client component. The server component is responsible for the communication with the smart luminaires, collecting and storing environment and location data, data analysis, generating reports and alerts, as well as sending generated alerts to responsible people. The client component is a web application that allows users to configure the system by providing the floor plans of monitored dwellings, registering new luminaires and system users, as well as to generate and view reports.

We briefly depict the main server subsystems. The \textit{indoor localization subsystem} uses RSSI data collected by smart and dummy luminaires and sent to the server. Several times each minute, this component calculates the indoor location of monitored users, and stores this information in the database, where it can be accessed for analysis. The \textit{data acquisition and analysis subsystem} receives data collected by the wireless network's luminaires and sends these to the persistence database. Furthermore, it also performs several types of analyses on the data to identify abnormal situations. The \textit{real-time alerts subsystem} is responsible for creating notifications and alerts if a situation that presents risk for the monitored person has been identified. Alerts are persisted in the database, and are sent in real time to both the monitored person and their caregivers via short message service. The \textit{reporting subsystem} is responsible for creating different types of reports, using data collected by the intelligent luminaires and stored in the database. The subsystem exposes a web service for providing statistics and reports. These can be visualised through the web application. The \textit{web application} is a web interface for configuring system preferences, managing intelligent luminaires, users and their associated locations, and viewing different types of reports.

\section{Indoor Localization Techniques}
\label{localisation_techniques}
In order to increase the accuracy of indoor localization, and achieve a reduction in positioning errors, we evaluate several approaches starting with direct trilateration, which is then improved using several proposed heuristics. An additional technique we evaluate is based on an artificial neural network trained to produce coordinates associated to the person's indoor location.

\subsection{Direct Trilateration}
\label{sec:trilateration}
The indoor localization algorithm developed for i-Light is based on trilateration. As opposed to triangulation, which employs angles, trilateration computes the position of a target starting from the computed distances between the target and at least three static receivers. In our case, intelligent luminaires are the receivers and a smart device worn by the monitored person is the target. To compute the required distances we start from RSSI measurements and use the RSSI lognormal model described in \cite{dong2012}. Equation \ref{eq:localisation} describes the manner of calculating the distance, using the RSSI value and 2 other parameters: $n$, the path-loss exponent and $A$, the signal strength expressed in dBm, measured at a distance of one meter. The path-loss exponent's value is directly influenced by the physical environment in which the system operates. This includes the surrounding structure, building materials and presence of furniture in the case of indoor environments. Its values range between 1.4 to 5.1, depending on the environment \cite{huang2015}. 

\begin{equation}
\label{eq:localisation}
d=10^{\frac{A-RSSI}{10 \cdot n}}
\end{equation}

\begin{figure}[h]
    \centering
    \includegraphics[width=0.6\linewidth]{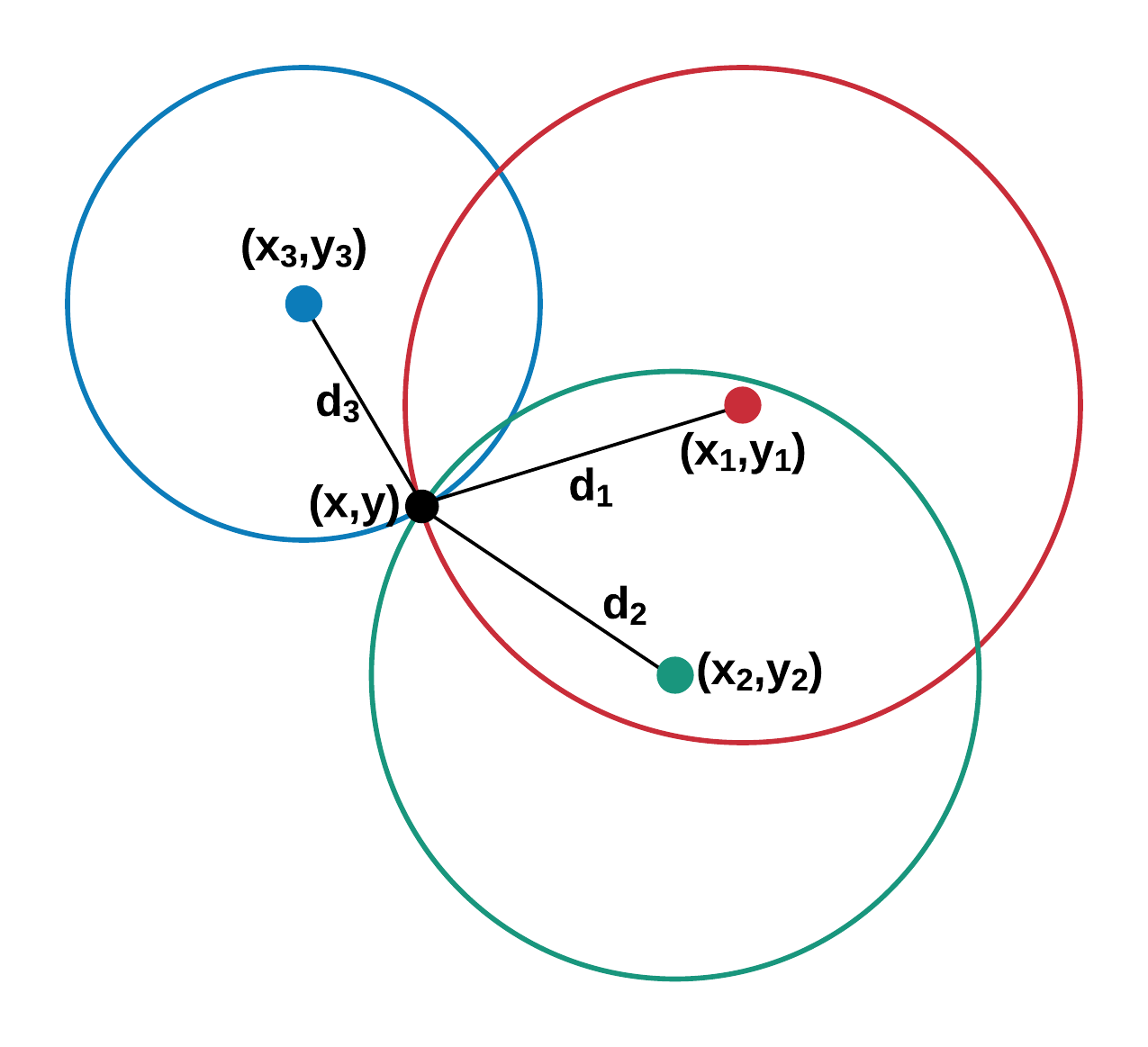}
    \caption{Illustration for trilateration. The $(x_{i}, y_{i}), i \in \{1, 2, 3\}$ dots represent luminaires and the $(x,y)$ dot is the computed position, considering the three circles' radii.}
    \label{fig:trilateration}
\end{figure}

Having three distances computed according to Formula \ref{eq:localisation}, from the target of monitoring to each of the installed luminaires, the position of the target within the environment can be inferred. This is computed as the point of intersection of three circles, each having as centre one luminaire and as radius the computed distance from that luminaire to the target, as shown in Figure \ref{fig:trilateration}.

The target's coordinates, denoted $(x, y)$, are obtained by solving the system of equations in Formula \ref{eq:trilateration}, where $(x_{i}, y_{i}), i \in \{1, 2, 3\}$ are the luminaires' positions (the circles' centres) and $d_i, i \in \{1, 2, 3\}$ represent the respective computed distances (the circle radii ($d_1, d_2, d_3$)).

\begin{equation}
\label{eq:trilateration}
  \systeme*{
  (x - x_{1})^2 + (y - y_{1})^2 = d_1^2,
  (x - x_{2})^2 + (y - y_{2})^2 = d_2^2,
  (x - x_{3})^2 + (y - y_{3})^2 = d_3^2
  }
\end{equation}

While the above-mentioned positioning method is theoretically sound, in real-world environments, RSSI values are attenuated by various types of obstacles, such as furniture or other indoor objects and walls \cite{brena2017}. Wall type and thickness has a direct influence on signal strength \cite{zargoun2016}, meaning that the equation system cannot be effectively solved. To handle signal noise the following subsections present a number of proposed methods that can be employed on the server side to improve localization accuracy.

\subsection{Kalman Filter}
The Kalman filter is an algorithm for optimal estimation in linear Gaussian systems \cite{de1999}. It is suitable for dynamic systems in which the goal is to find the best estimate of a state from indirect or noisy measurements. The technique is well-suited for real-time application, as in order to estimate the current state it only needs information from the previous one. Taking into consideration the technique's purview and advantages, as well as our objectives regarding indoor localization, we decided to select this method for experimentation.

Indoor objects and walls directly influence the received signal strength \cite{brena2017}, thus the measurements collected by the intelligent luminaires at discrete points in time are noisy. To obtain the best estimate for the actual RSSI values at one point in time, Kalman filters use a prediction from the best estimate of the previous point in time along with a correction for known external factors, all starting from the assumption that the system variables are random and Gaussian distributed. Starting from some measured values, the algorithm proceeds with two major steps: \textit{prediction} and \textit{update}. First it makes a prediction of the current state, based on the previous state and problem model and then it updates this prediction, based on values measured at that time point and considering potential errors. Another aspect worth mentioning is the \textit{Kalman gain}, which provides more weight to either the estimate or the measurement, according to prediction error. A decrease in estimation error will result in more weight given to estimates. The estimated RSSI values obtained through Kalman filtering will further be used for trilateration, as described in subsection \ref{sec:trilateration}.

\subsection{Look-back-k Heuristic}
We propose and evaluate a second technique for error reduction. It is a generalization of the heuristic first presented in \cite{marin2018}. Its name is representative for its main feature: it considers the \textit{k} previous consecutive measurements for the computation of the current location. The underlying assumption is that a set of several measurements are more accurate than the last one and that by examining several measurements instead of just the current one, the precision can be increased.

In direct trilateration the three distances are computed at each discrete time point when an RSSI measurement is taken, using as input the currently measured RSSI value. If one measurement is faulty, it negatively influences the accuracy of the calculated position. We attempt to alleviate this issue by considering a collection of previous $n$ measurements. For each luminaire $l_i$, a series of RSSI values ($v$) are collected at $m$ time points $\{t_1, t_2, \dots, t_m\}$, as follows: $(v_{l_i}^{t_1}, v_{l_i}^{t_2}, \dots, v_{l_i}^{t_m})$ where $v_{l_i}^{t_j}$ represents the RSSI value for luminaire $l_i$, $i \in \{1, 2, 3\}$, at time point $t_j$, $j \in \{1, \dots, m\}$. Instead of computing the indoor position $p_{t_j}$ at time point $t_j$ by employing the three values $(v_{l_1}^{t_j}, v_{l_3}^{t_j}, v_{l_3}^{t_j})$, we start from the previous $k$ RSSI measurements, for each luminaire $\{v_{l_i}^{t_{j-(k-1)}}, \dots, v_{l_i}^{t_{j-2}}, v_{l_i}^{t_{j-1}}, v_{l_i}^{t_j}\}$ and perform the following computations for time point $t_j$, on this collection: 

\begin{enumerate}
    \item Eliminate outliers. We take into account two procedures for outliers: (1) The minimum and maximum values are outliers; (2) We consider a value to be an outlier if it falls outside of 1.5 times of an interquartile range above the third quartile and below the first quartile.
    \item Compute the mean $\mu_{t_j}$ and the standard deviation $\sigma_{t_j}$ for the remaining values.
    \item Eliminate all values that are further than one standard deviation from the mean, thus all values outside the interval $[\mu_{t_j} - \sigma_{t_j}, \mu_{t_j} + \sigma_{t_j}]$.
    \item Compute the mean of the remaining values $\mu'_{t_j}$ and use it in the trilateration process to compute the distance.
\end{enumerate}

\subsection{Hybrid Technique}
Previously described techniques are intended to reduce localization errors resulting from noisy or erroneous measurements or as a result of interference. A combination of the two could steer towards more accurate indoor positioning results. Hence, this hybrid technique works in two phases: first, RSSI estimates are computed using the Kalman filter after which the look-back-n technique is applied using the estimates as input data.

\subsection{Neural Network Based Technique}
\label{sec:ann}
While the first three techniques rely mainly on trilateration and heuristic methods applied to minimise the impact of noise on input data, this technique stems from machine learning, where it is ubiquitous for both classification and regression. Several recent studies tackling indoor positioning have employed such machine learning models: a four layer deep neural network, combined with a denoising autoencoder and a hidden Markov model is employed in \cite{zhang2016} for indoor and outdoor localization; a recurrent neural network which uses WiFi signals for an indoor positioning system is presented in \cite{lukito2017}; another recurrent neural network, more specifically a long short-term memory network is used by Urano et al. \cite{urano2019} with BLE signal strength data for indoor localization; Mittal et al. \cite{mittal2018} propose a convolutional neural networks based framework for indoor localization, in which the networks use images created from WiFi signatures. Another convolutional deep neural network starting from phase data of channel state information, which is transformed into images based on estimated angles of arrival is presented in \cite{wang2018}. In our model, the neural network input is a triplet that contains the three RSSI values recorded by the intelligent luminaires, while the output is the estimated location, represented as a coordinate pair.

Some of the most important hyper-parameters to fine-tune when working with an artificial neural network are the number of layers, the number of neurons per each hidden layer, type of optimisation algorithm, types of activation functions, learning rate or regularisation.  However, there is no universal solution for deciding the optimal parameters' values for a given problem description (and non-linearly separable data) and most network architectures are built based on prior experience, trial and error. Our network starts from a simple architecture, with only one hidden layer. To this we gradually add hidden layers to investigate how this influences the obtained accuracy and to contribute to improved localization. The activation function of choice is rectified linear unit (ReLU) \cite{nair2010} and the optimisation algorithm employed is \emph{adam} \cite{adam}, an extension to stochastic gradient descent often used in the field of deep learning. Stratified k-fold cross validation is used during training for a more robust model, while the number of training epochs is varied during experimentation.

\section{Experimental Evaluation}
\subsection{Methodology}
\label{methodology}
Given the known impact building layouts and materials have on wireless signal propagation, we carried out our evaluation in two distinct locations. To allow for directly comparing the achieved accuracy, the same three intelligent luminaires were employed in both cases. 

\begin{figure*}[h]
  \subfloat[Home \cite{marin19}]{\label{fig:dwelling_home}
      \includegraphics[width=.48\textwidth]{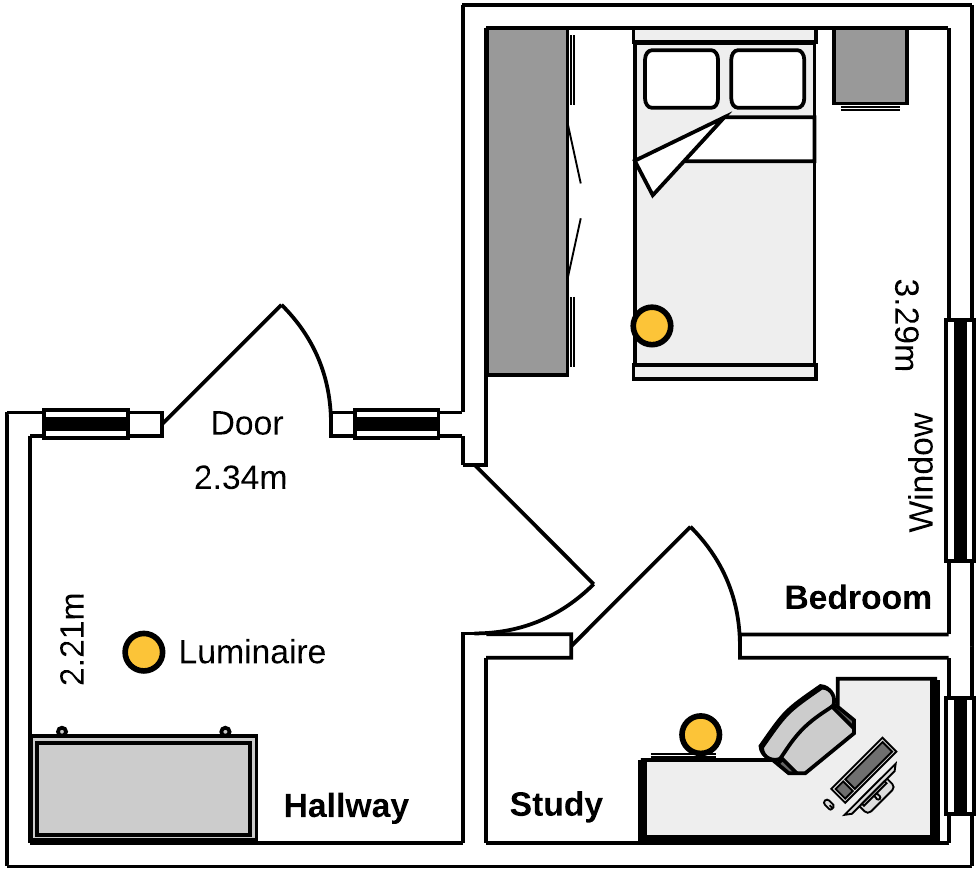}}
~
  \subfloat[Office]{\label{fig:dwelling_office}
      \includegraphics[width=.48\textwidth]{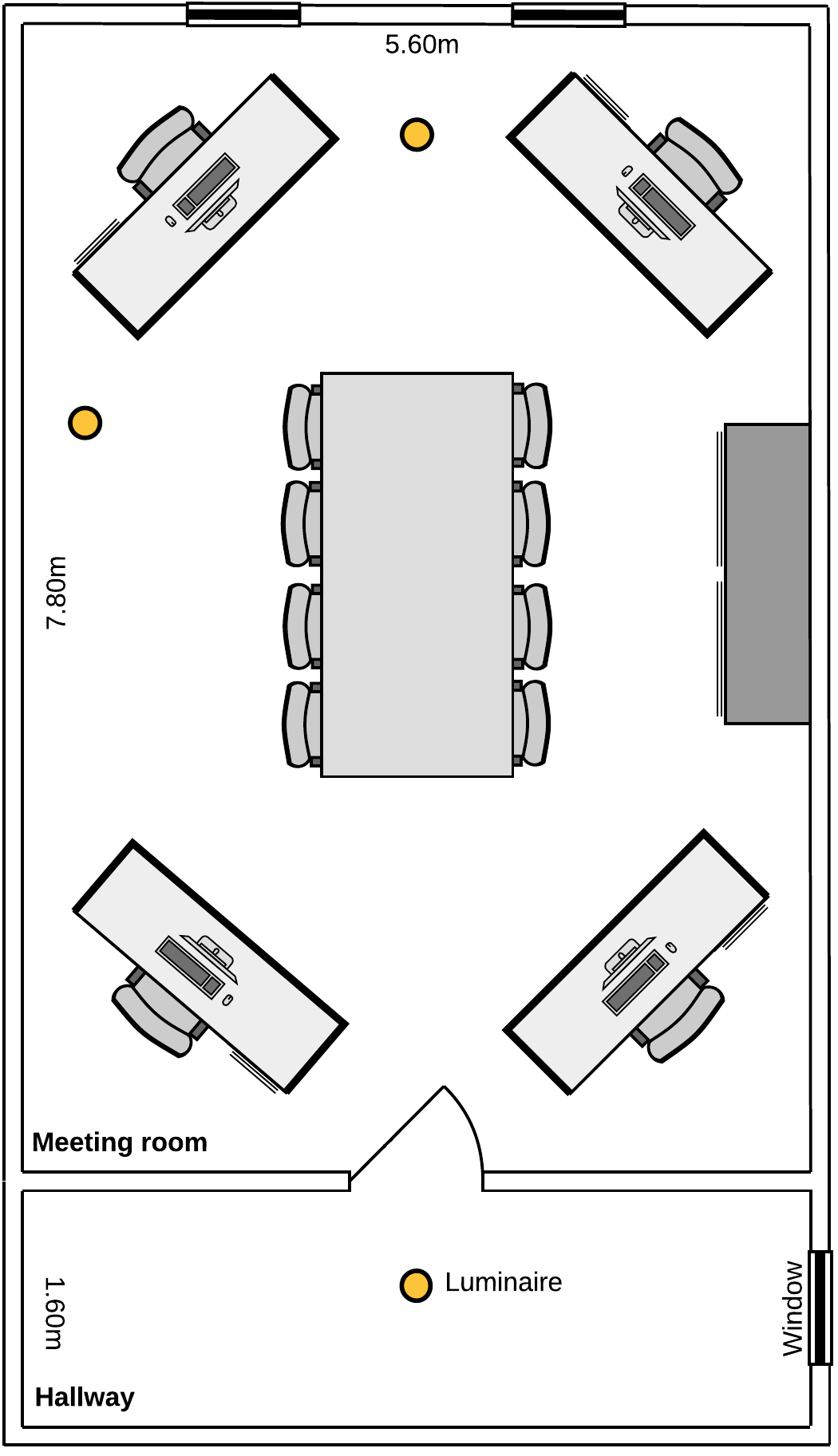}}

  \caption{Partial floor map of dwellings used for evaluation (drawings to scale).}
  \label{fig:dwelling}
\end{figure*}

The first location was a home, having room dimensions of 2.50m x 3.29m (bedroom), 2.50m x 1.00m (study room), 2.34m x 2.21m (hallway). Additional rooms, such as the kitchen and bathroom were not included in our evaluation. Figure \ref{fig:dwelling_home} illustrates the layout of the evaluated rooms. One luminaire was ceiling-mounted in each room, replacing existing lighting infrastructure. Luminaire positions in Figure \ref{fig:dwelling} are marked using yellow circles with black outlines. 

The second location was a two-room section of a large multi-storey office building, illustrated in Figure \ref{fig:dwelling_office}. Enclosure sizes are 5.60m x 7.80m for the meeting room and 1.60m x 5.60m for the hallway. The office covers more than 3 times the home's floor space, as shown in Figure \ref{fig:dwelling}, which is drawn to scale. Two luminaires were deployed in the meeting room, while the third one was installed in the hallway. Both figures include major pieces of furniture, which have a detrimental effect on signal transmission \cite{wang2012}. More importantly, building materials used differ across the considered locations. The home location has both interior and exterior walls of brick and cement, with interior walls being 17cm thick, while exterior ones are 35cm \cite{marin19}. For the office location, exterior walls were made of autoclaved aerated concrete and the indoor ones made of plasterboard. It is known that aerated concrete absorbs electromagnetic waves \cite{laukaitis2008}. The electromagnetic shielding of the plasterboards is proportionally greater as the fiber content increases, while with increasing environment moisture, the shielding effect is decreased \cite{samkova2018}. Signal interference was significant within the office location scenario, with 14 WiFi systems and 11 Bluetooth devices enabled around the testing area creating impedance. In both locations, a smartphone located 1m above the floor was used as signal emitter, while the luminaires acted as signal receivers.

The trilateration formula presented in Section \ref{sec:trilateration} employs two parameters that need to be calibrated: $A$ - the signal strength at 1 meter from the luminaire and $n$ - the path loss exponent. As determined in our experiments the value for $A$ is different for the two scenarios ($-87$ for the first experiment and $-67$ for the second). As reported in existing literature \cite{miranda13,okorogu13}, the value of $n$ varies depending on room shapes and sizes, building materials, wall placement as well as furniture. This value was determined by a grid-search procedure, showing that the most suitable value for both scenarios is $n=2.5$.

The experimental methodology was similar in both locations. First, the devices were installed and we ensured that a stable link to the remotely-deployed server was working. Then, one person playing the role of \textit{monitored adult} was stationary in several fixed places within the monitored rooms for a duration of around 15 minutes. A mobile phone was kept on them at a height of 1m and with Bluetooth turned on, but not paired to another device. The luminaires recorded RSSI values from the phone at under 10 second intervals each, and sent the raw data to the server for further analysis. The following section details the evaluation carried out on the raw data. Post processing was done exclusively on the software server. The presented methods were implemented server-side and can run in real-time under deployment conditions.

\subsection{Results}
We present below the results obtained after applying the techniques presented in Section \ref{localisation_techniques} for indoor localization, using the two scenarios presented in Section \ref{methodology}. The \textit{Euclidean distance} was selected as the evaluation measure and the obtained distances were averaged across time points, for each interval. It must be mentioned that RSSI timestamps were not exactly identical for all the rooms in which the measurements were taken. In order to compute the person's location in any given moment \textit{t}, the closest timestamps to moment \textit{t}, for each room, were used.

\subsubsection{Home location.} 
\begin{figure*}[htp]
   \subfloat[House - person is in the bedroom.]{\label{fig:interval1}
      \fbox{\includegraphics[width=.48\textwidth]{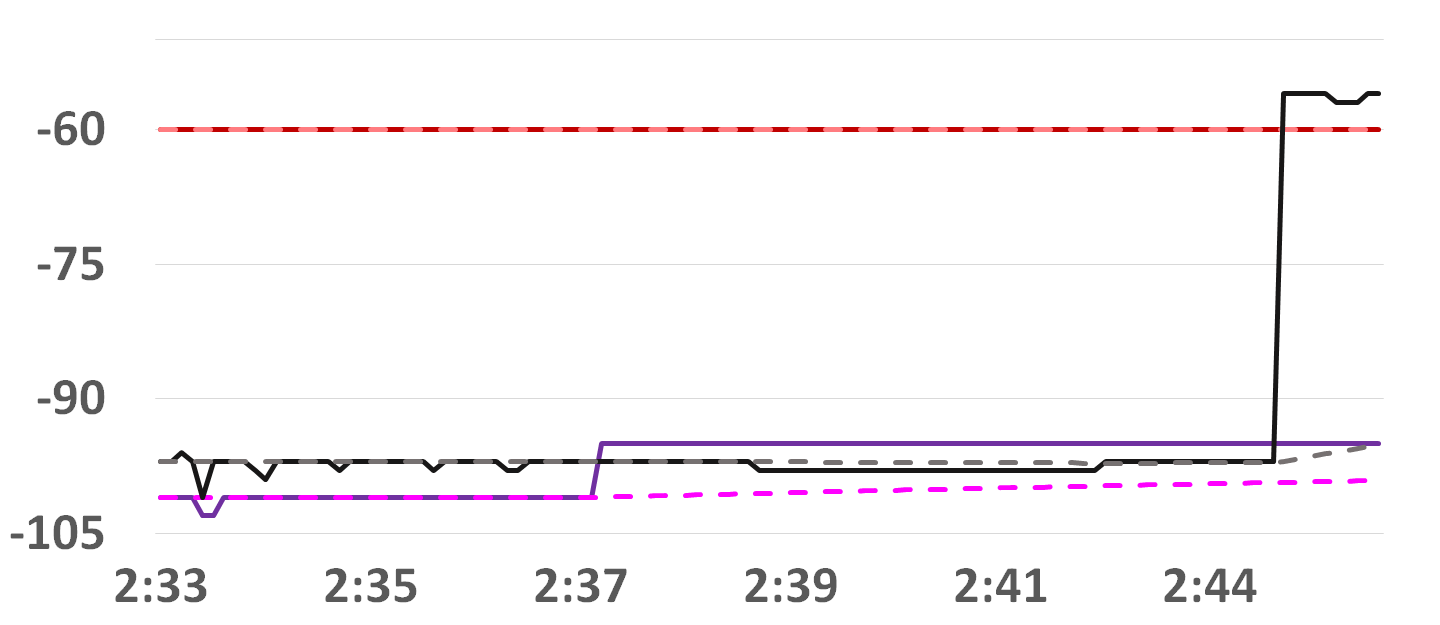}}}
~
   \subfloat[House - person is in the study room.]{\label{fig:interval2}
      \fbox{\includegraphics[width=.48\textwidth]{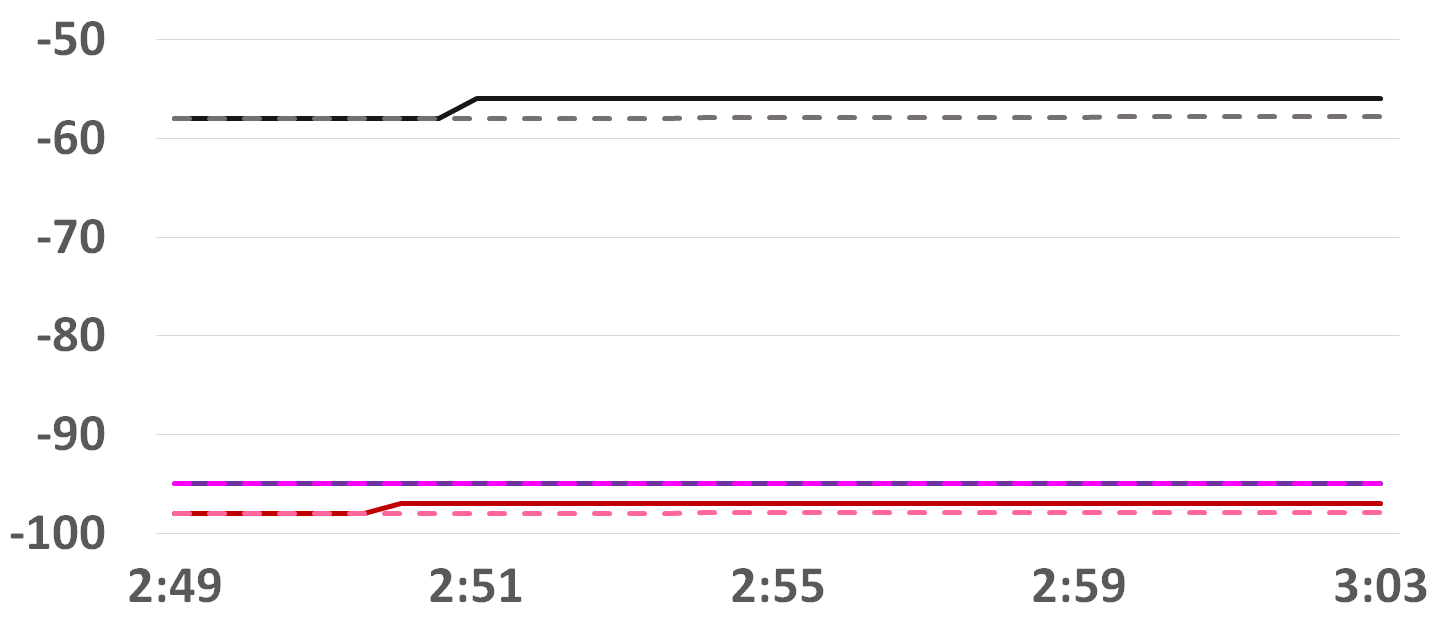}}}
 
   \subfloat[House - person is in the hallway.]{\label{fig:interval3}
      \fbox{\includegraphics[width=.48\textwidth]{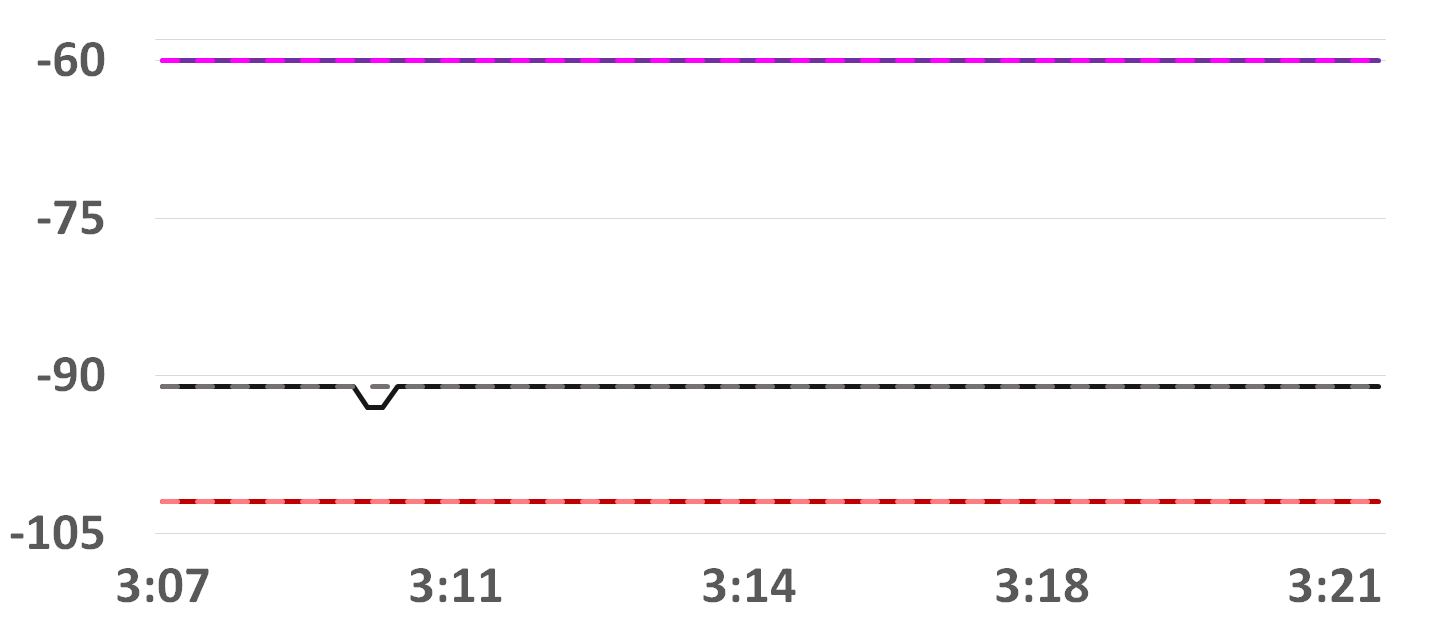}}}
~   
   \subfloat[Office - person is in meeting room.]{\label{fig:interval4}
       \fbox{\includegraphics[width=.48\textwidth]{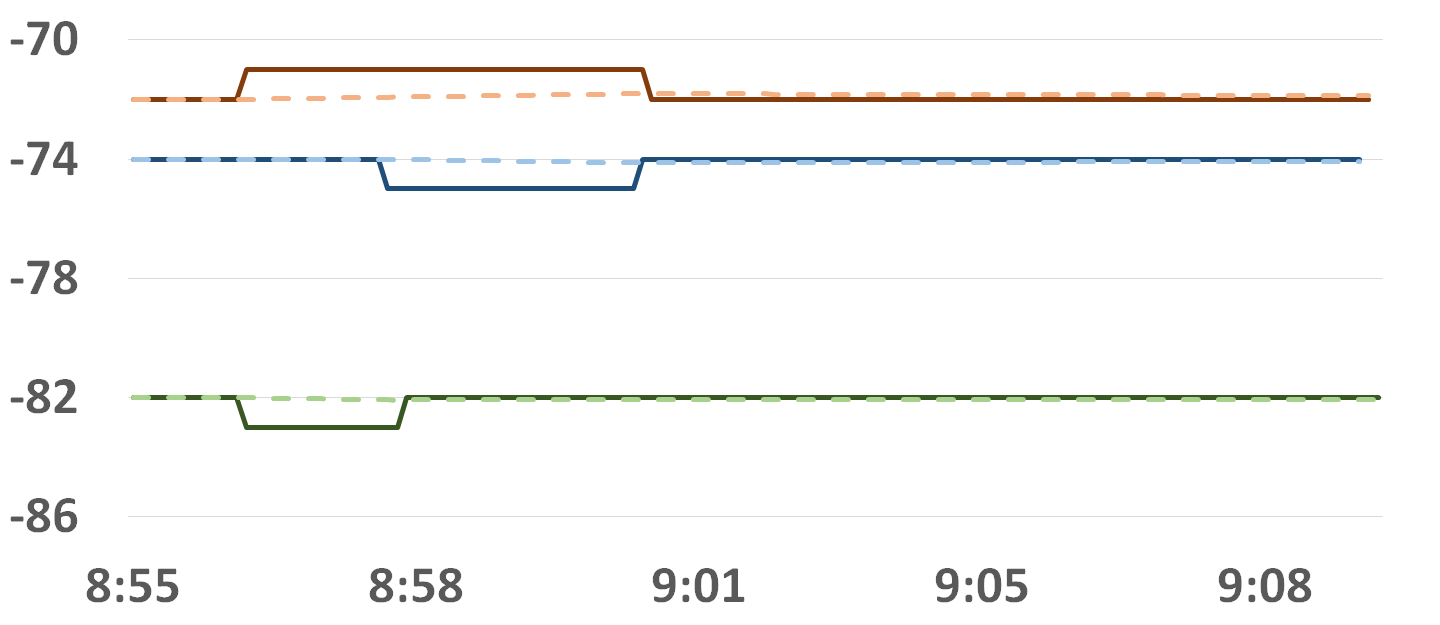}}}
   
   \subfloat[Office - person is in meeting room.]{\label{fig:interval5}
       \fbox{\includegraphics[width=.48\textwidth]{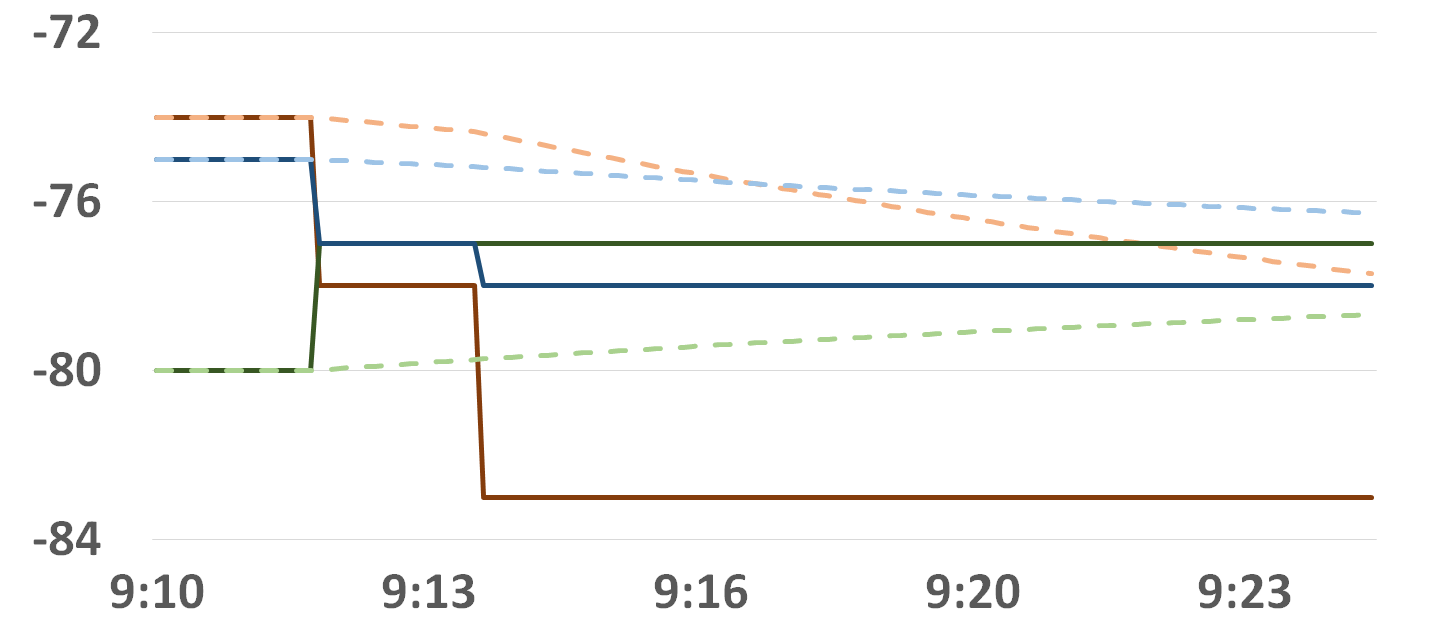}}}
~
   \subfloat[Office - person is in meeting room.]{\label{fig:interval6}
       \fbox{\includegraphics[width=.48\textwidth]{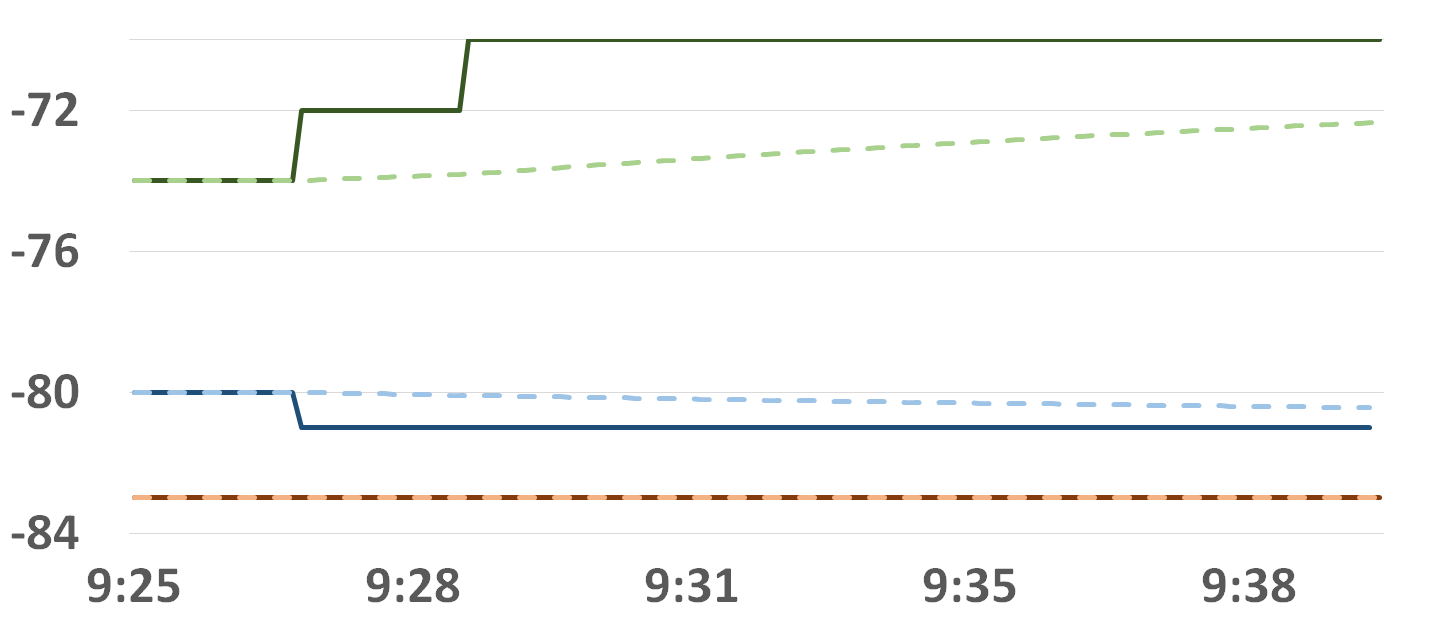}}}

   \subfloat[Office - person is in hallway.]{\label{fig:interval7}
       \fbox{\includegraphics[width=.48\textwidth]{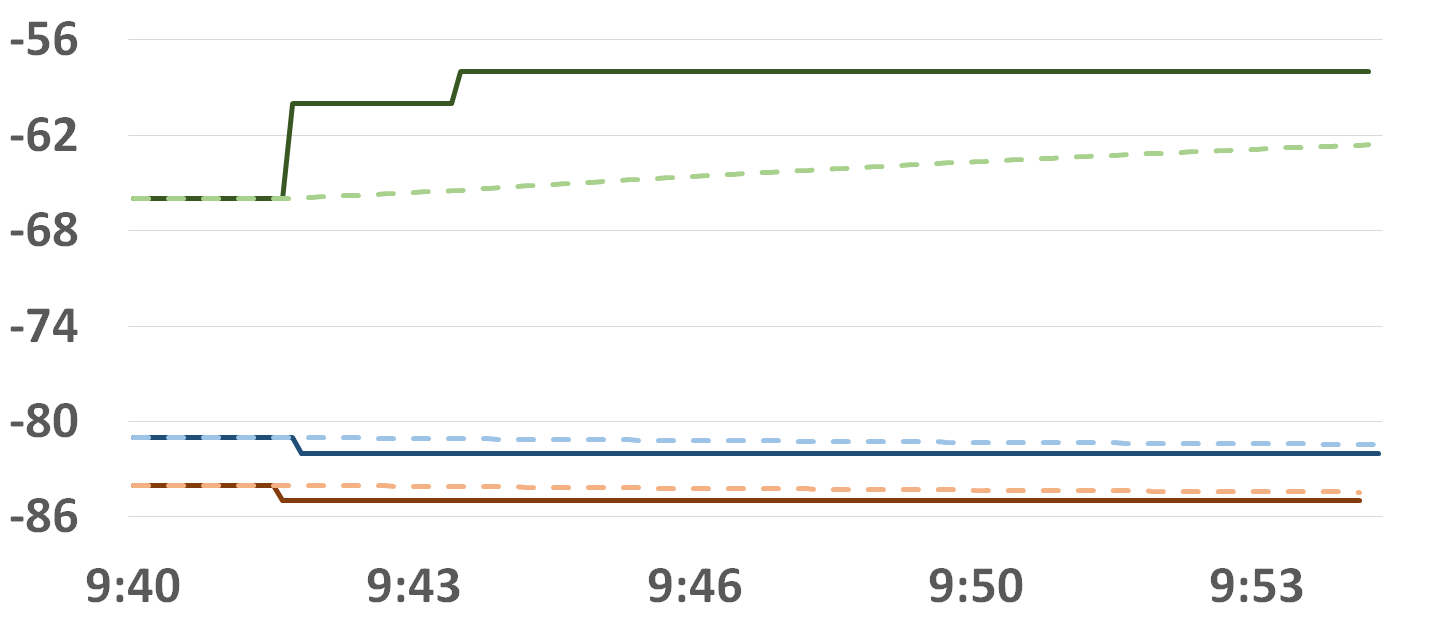}}}
~
   \subfloat[Office - person is in hallway.]{\label{fig:interval8}
       \fbox{\includegraphics[width=.48\textwidth]{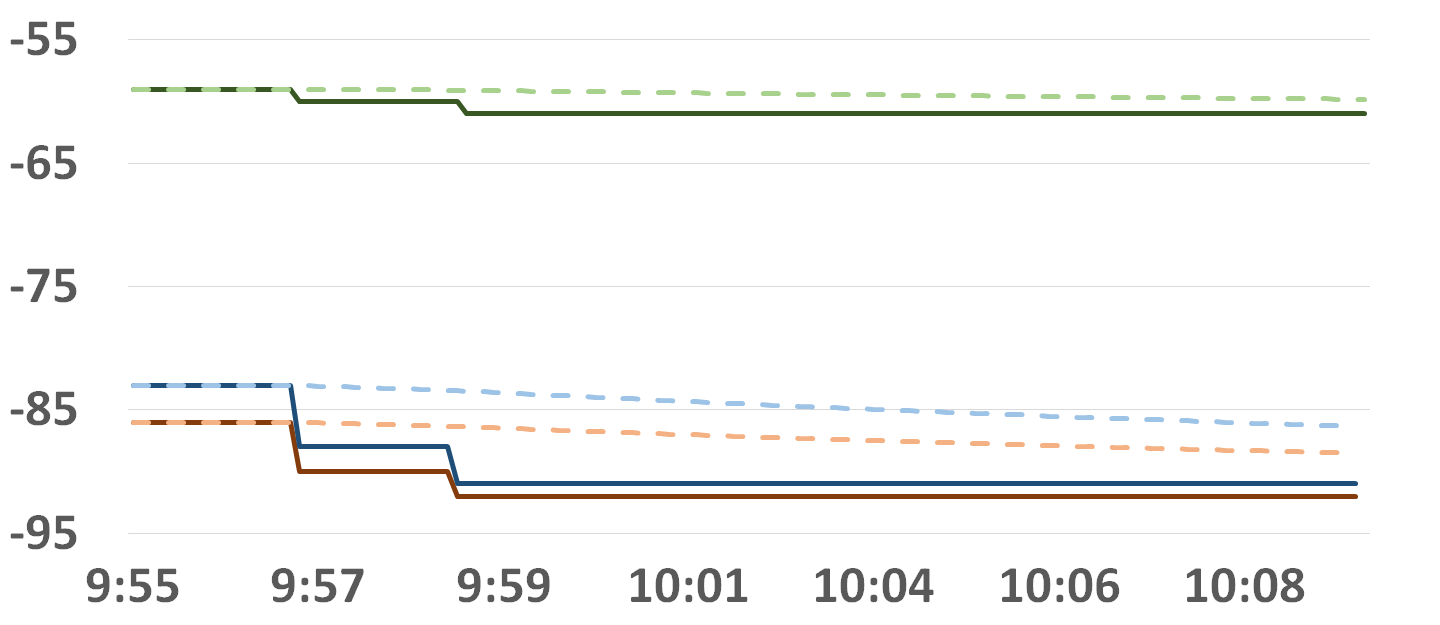}}}

   \caption{Raw RSSIs (continuous lines) and Kalman estimates (dashed lines) for each time interval. Signal strength ({dBm}) on vertical axis, time on horizontal axis. Luminaires are color coded as follows: magenta (house hallway), black (study room), red (bedroom), orange (meeting room, by the window), blue (meeting room by the wall), green (office hallway).}
   \label{fig:kalman}
\end{figure*}

The first location for evaluation is the three-room home. Figures \ref{fig:interval1}, \ref{fig:interval2} and \ref{fig:interval3} provide a visual representation of the evolution of RSSI values, plotted against the corresponding Kalman estimates. Inspecting these values we notice that during the last two time periods, when the person was in the study room and hallway, the values are inconsistent during the time interval. However, RSSI values recorded by the study bulb while the person was in the bedroom show a significant rise towards the end of the period, which is inconsistent with the other values and thus suggests noisy measurements. This jump is significantly softened using Kalman filtering.

Results obtained using all four direct trilateration-based techniques described in Section \ref{localisation_techniques} are presented in Table \ref{table:first_experiment}. Each column corresponds to the period the monitored person spent in each room and the last column shows the average error. For the look-back-k technique we consider several values for the parameter $k$ and show results for all of them. A first observation is that the look-back-k heuristic, in which a set of values is used for position computation is on average more efficient than using time point characteristic values. For the first time period, when the measurements seem to be noisier, especially towards the end of the period, we observe a general tendency of decreasing error with the increase of $k$. For the other two periods, the error either slightly increases or remains the same. However the average positioning error is approximately $10$cm lower for $k=50$, when compared with using raw RSSI values. The reported results correspond to the minimum and maximum-based outlier elimination. The interquartile range-based elimination yields highly similar results: for $k=50$, the returned error is $126.36$cm (as opposed to $125.80$cm, with the minimum and maximum-based elimination). The same decreasing trend is observed in the case of interquartile range-based elimination.

Kalman filters bring additional improvements. When applying these filters on raw values, and further, using trilateration starting from Kalman estimates, the positioning error is even smaller: more than 21 centimetres are gained when compared to using raw values and approximately 12 centimetres when compared to the best result using the look-back technique. Considering the improvements brought by both proposed heuristics, the expectation is that combining them will improve localization even further. This is confirmed by our experiment, as the best result obtained by the hybrid technique leads to the smallest error. The largest difference in error is for the first time interval (column 2:33-2:48), which is expected when considering the noisy values recorded towards the end of the interval, as shown in Figure \ref{fig:kalman}. Analysing the results of this experiment we conclude that the noise reduction induced by Kalman filtering brings significant enhancement for indoor localization, while the combination of this technique with our proposed look-back-k heuristic can lead to additional improvements.

\begin{table}[t]
\footnotesize
\centering
\begin{tabular}{ c c|c|c|c| } 
\cline{2-5} & \multicolumn{1}{|c|}{2:33-2:48} & 2:49-3:04 & 3:07-3:22 & Avg. error \\ 
\hline
\multicolumn{1}{|c|}{Raw values}  & 141.67  &  30.71  & 233.18   & 135.19 \\ 
\hline
\multicolumn{1}{|c|}{Look-back-5} & 137.41  & 31.98    & 233.23  & 134.21 \\ 
\hline
\multicolumn{1}{|c|}{Look-back-10} & 134.22  &  33.07   &  233.24  & 133.51 \\ 
\hline
\multicolumn{1}{|c|}{Look-back-15} & 131.39 &  33.82   &  233.25  & 132.82 \\
\hline
\multicolumn{1}{|c|}{Look-back-20} & 128.63  &  34.43   &  233.24  & 132.1 \\
\hline
\multicolumn{1}{|c|}{Look-back-30} & 122.51  &  34.94  &  233.17  & 130.21 \\
\hline
\multicolumn{1}{|c|}{Look-back-50} & 109.40  &  34.94   & 233.05   & 125.80 \\
\hline
\multicolumn{1}{|c|}{Kalman filter} & 64.88  & 43.16  & 233.56 & 113.87 \\ 
\hline
\multicolumn{1}{|c|}{Kalman filter +} & \multirow{2}{*}{63.92}  &  \multirow{2}{*}{43.32}  &  \multirow{2}{*}{233.56}  & \multirow{2}{*}{113.60} \\ 
\multicolumn{1}{|c|}{look-back-5} &   &    &    &  \\ 
\hline
\multicolumn{1}{|c|}{Kalman filter +} & \multirow{2}{*}{62.79}  &  \multirow{2}{*}{43.48}  &  \multirow{2}{*}{233.56}  & \multirow{2}{*}{113.28} \\ 
\multicolumn{1}{|c|}{look-back-10} &   &    &    &  \\ 
\hline
\multicolumn{1}{|c|}{Kalman filter +} & \multirow{2}{*}{61.72}  &  \multirow{2}{*}{43.59}  &  \multirow{2}{*}{233.55}  & \multirow{2}{*}{112.95} \\ 
\multicolumn{1}{|c|}{look-back-15} &   &    &    &  \\ 
\hline
\multicolumn{1}{|c|}{Kalman filter +} & \multirow{2}{*}{60.72}  &  \multirow{2}{*}{43.65}  &  \multirow{2}{*}{233.55}  & \multirow{2}{*}{112.64} \\ 
\multicolumn{1}{|c|}{look-back-20} &   &    &    &  \\ 
\hline
\multicolumn{1}{|c|}{Kalman filter +} & \multirow{2}{*}{58.79}  &  \multirow{2}{*}{43.67}  &  \multirow{2}{*}{233.55}  & \multirow{2}{*}{112.00} \\ 
\multicolumn{1}{|c|}{look-back-30} &   &    &    &  \\ 
\hline
\multicolumn{1}{|c|}{Kalman filter +} & \multirow{2}{*}{55.38}  &  \multirow{2}{*}{43.67}  &  \multirow{2}{*}{233.55}  & \multirow{2}{*}{110.87} \\ 
\multicolumn{1}{|c|}{look-back-50} &   &    &    &  \\ 
\hline
\end{tabular}
\caption{Localization errors for home location. Errors computed as averaged Euclidean distances, reported in centimetres.}
\label{table:first_experiment}
\end{table}

Raw RSSI values were fed to the neural network described in Section \ref{sec:ann}, whose architecture was updated in an iterative manner to investigate the changes in localization accuracy induced by network construction. Table \ref{table:first_experiment_ann} presents the results obtained for different numbers of hidden layers and increasingly more training epochs. All hidden layers have the same number of neurons, which was determined through experimentation. The network's performance was measured using a stratified 10-fold cross validation technique. Reported results are the averaged errors and standard deviations, based on Euclidean distances, obtained over 10 runs, for all three time periods. More training epochs generally induce lower errors, especially when passing from a lower number (100) to considerably higher ones (500, 1000 and so on). However, as soon as convergence is achieved, a further increase in the number of epochs does not bring such significant improvement, as can be seen in the case of 1000, 2000 and 3000 epochs. Alteration of the number of hidden layers causes significant changes in accuracy particularly for networks that have been trained over few epochs, as can be seen for 100 epochs, where the minimum error, obtained for 5 layers is approximately 35 centimetres less than the maximum one, obtained for just one hidden layer. Again, when the number of training epochs is high, such significant differences no longer appear. Analysing the results in Table \ref{table:first_experiment_ann} we observe that accuracy less than 10 centimetres is obtained for 1000 epochs and fewer hidden layers.

For the conducted experiments, the highest average error obtained by the network is 48.13cm, obtained for  100 training epochs and 1 hidden layer. This indicates that the neural network is clearly superior to the methods based on trilateration, where the lowest error obtained after applying the proposed heuristics is 110.87cm in the case of this experiment. The network obtains almost perfect performance in certain configurations, such as an error of 1.41cm for 3000 training epochs and 3 hidden layers. On average, the network's performance is altogether efficient, as the average error for the conducted experiments is 10.51cm and the $95\%$ confidence interval is $10.51 \pm 4.02$cm ($[6.49, 14.53]$). While the network is trained for three fixed positions, a network constructed for real time localization should be trained using data collected from multiple positions that cover the entire indoor area and in various settings. Thus, although the artificial neural network seems to be more advantageous than direct trilateration, it also presents some drawbacks, which might make it more difficult to employ during product deployment.

\begin{table}[t]
\footnotesize
\centering
\begin{tabular}{c c|c|c|c|c|}
\cline{2-6} & \multicolumn{5}{|c|}{Number of hidden layers} \\
\hline
\multicolumn{1}{|c|}{Epochs} & 1 & 2 & 3 & 4 & 5 \\ 
\hline
\multicolumn{1}{|c|}{100}  & $48.13 \pm 37.24$ & $ 25.98 \pm 10.27$  & $21.96 \pm 7.88$ & $ 24.76 \pm 21.72$ & $13.02 \pm 4.38$ \\ 
\hline
\multicolumn{1}{|c|}{500}  & $12.04 \pm 3.80$ & $10.92 \pm 3.48$  & $9.74 \pm 3.23$ & $6.04 \pm 2.66$ & $5.81 \pm 4.58$ \\ 
\hline
\multicolumn{1}{|c|}{1000} & $9.74 \pm 2.56$ & $7.85 \pm 4.22$  & $7.35 \pm 4.71$ & $7.47 \pm 3.39$ & $5.12 \pm 3.10$ \\ 
\hline
\multicolumn{1}{|c|}{2000} & $9.59 \pm 4.37$ & $6.28 \pm 3.57$ & $4.15 \pm 3.77$ & $2.94 \pm 2.24$ & $2.71 \pm 1.67$ \\ 
\hline
\multicolumn{1}{|c|}{3000} & $10.39 \pm 3.63$ & $5.20 \pm 4.26$  &1.41 $ \pm 0.87$  &  $2.69 \pm 2.26$ & $1.57 \pm 0.49$ \\ 
\hline
\end{tabular}
\caption{Home location. Average error based on Euclidean distance $\pm$ standard deviation obtained using the artificial neural network, over 10 runs, with stratified cross-validation. Errors reported in centimetres.}
\label{table:first_experiment_ann}
\end{table}

\begin{figure*}[h]
   \subfloat[Home]{\label{fig:loc_dwelling_home}
      \includegraphics[width=.48\textwidth]{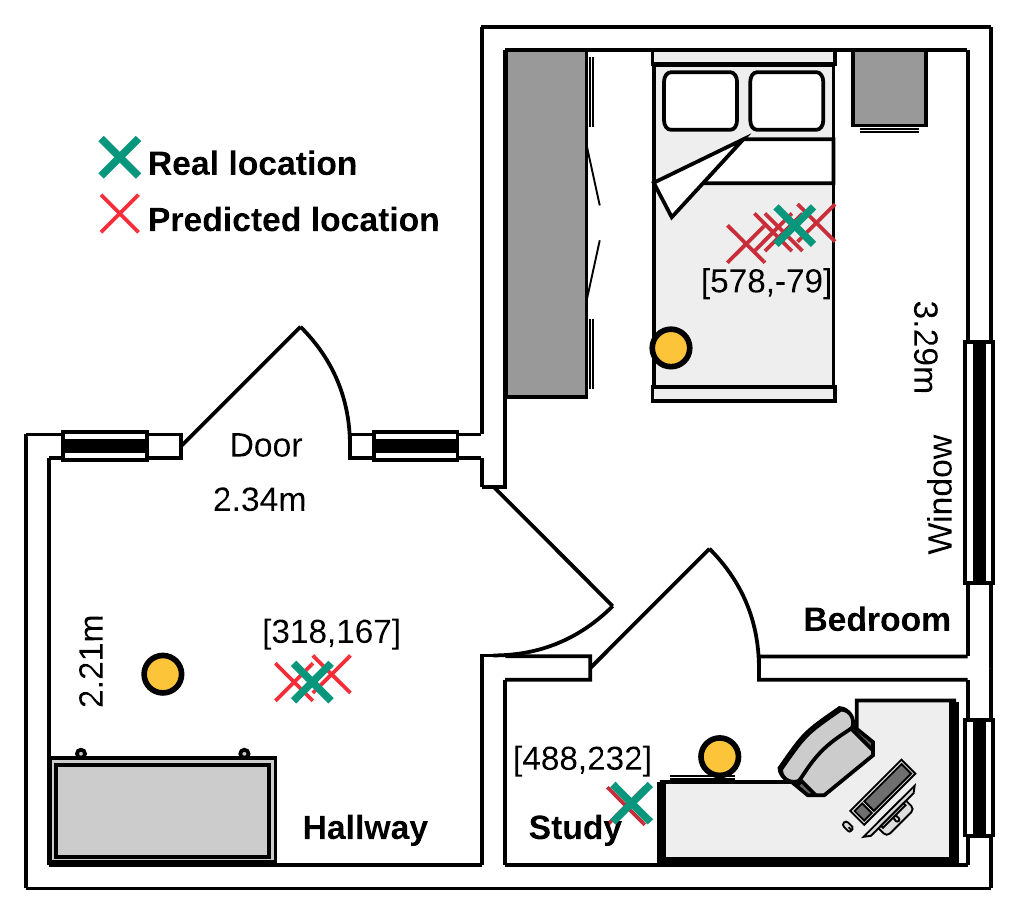}}
~
   \subfloat[Office]{\label{fig:loc_dwelling_office}
      \includegraphics[width=.48\textwidth]{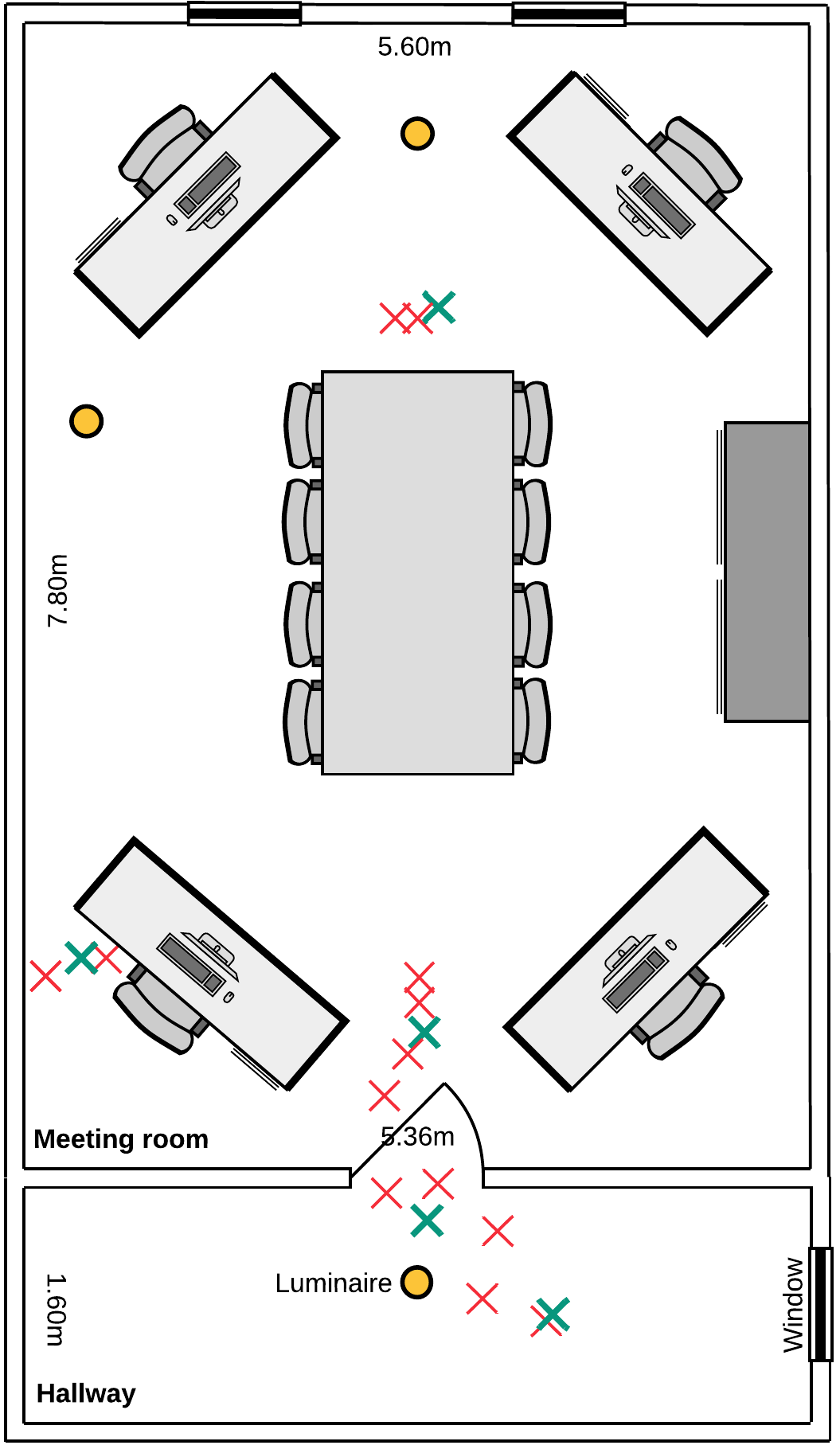}}

   \caption{Indoor positions obtained by the neural networks (3 hidden layers - left, 5 hidden layers - right), after being trained for 3000 epochs, represented on the two-dimensional floor plan of the dwelling (Figure \ref{fig:dwelling})}
   \label{fig:localization_network}
\end{figure*}

The actual and estimated locations obtained using the neural network are illustrated in Figure \ref{fig:localization_network}. Reported locations present the 5 hidden layers network configuration, which performed very accurately, with the computed positions obtained by the network plotted against the actual locations. As seen in Figure \ref{fig:loc_dwelling_home}, predicted positions are very close and in many cases overlap the actual location of the monitored person.

\begin{figure}[h!]
\includegraphics[width=\linewidth]{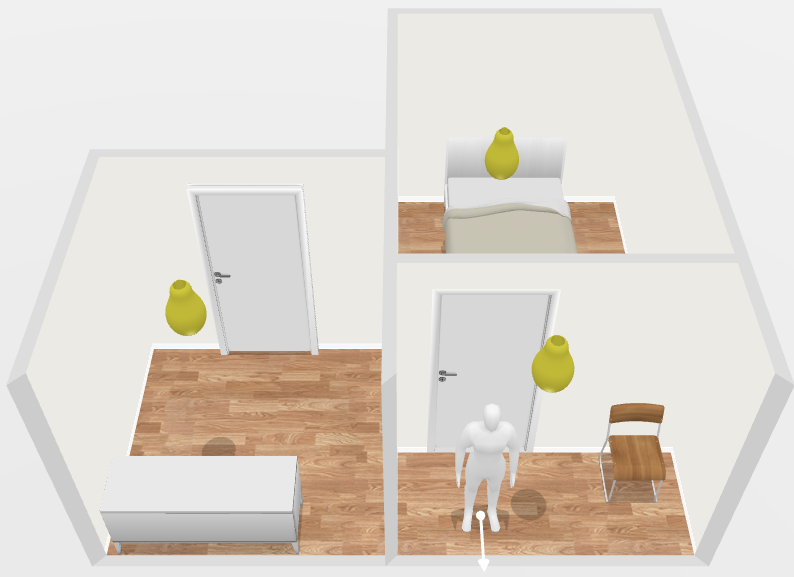}
\caption{Illustration of the person's indoor location within the system's interface \cite{marin19}.}
\label{fig:localisation_3D}
\end{figure}

Figure \ref{fig:localisation_3D} illustrates the location of the person during the second time interval, as it is reported by the cyber-physical system's software application. It allows users to map and display the dwelling floor plan together with the estimated real-time location of the monitored person using both 2D and 3D representations.

\subsubsection{Office location.}
In this section we present the results obtained during the evaluation carried out in the office location. In order to facilitate comparison with results obtained in the previous evaluation, the same methodology was followed. RSSI values were collected from five positions within the location illustrated in Figure \ref{fig:dwelling_office}. Recorded RSSI values and their Kalman estimates are illustrated in Figures \ref{fig:interval4}-\ref{fig:interval8}. For the first three intervals of time (8:55 - 9:10, 9:10 - 9:25 and 9:25 - 9:40) the person was in different spots in the meting room. While for the first and third intervals the RSSIs measured by the three luminaires did not change notably, we noticed that during the second interval values recorded by the window bulb had a larger variance. The last two intervals of time, during which the person stood in the hallway, also presented some more significant changes in the recorded RSSI values. We observe that in all cases Kalman filtering attenuates the differences.

Table \ref{table:second_experiment} shows the localization errors for this experiment, for each interval of time and each technique described in Section \ref{localisation_techniques}. The obtained results are rather consistent with those obtained during the home scenario: the look-back-n technique and Kalman filtering seem to improve localization on average, but not in all individual cases. For instance, for the second interval of time we notice that the error obtained by processing the Kalman estimates is far greater that the one obtained by processing raw values. We posit this occurs due to Kalman estimates being more similar to the initially recorded RSSI values, which correspond to the person moving from one location to the other. Thus, even though for the best part of the interval the recorded values were lower (-83, for more than 11 out of 15 minutes), the Kalman estimates are closer to the initial values ([-74, -77]). This type of error can be attenuated by increasing the weight given to measurements recorded for longer periods. Like in the case of the previous evaluation, reported results correspond to the minimum and maximum-based outlier elimination. Again, the interquartile range-based elimination obtained similar results, but slightly higher errors ($232.13$cm for look-back-50, as opposed to $222.77$cm and $227.77$cm for Kalman filtering combined with look-back-50).

Using Kalman filters brings additional improvements, similar to the case of the house scenario. When the filter is applied on raw values, and further, when employing trilateration starting from the Kalman estimates, positioning errors are decreased, as shown in Table \ref{table:second_experiment}. Analysing the results presented in the table, we conclude that the Kalman filtering noise reduction again improves the accuracy of indoor localization, while its combination with our proposed look-back-k heuristic leads to additional improvements.

One of the differences between the house and office scenarios regards the accuracy of the proposed hybrid technique. While in the first scenario, the hybrid technique produced the smallest errors, in this case we observe different results. The best results with the hybrid technique are obtained when considering Kalman estimates and 5 steps back. All average errors output by Kalman filtering and the look-back-n technique are lower than those generated by using raw values. We observe that the results of this experiment are strongly influenced by the selected positions in the office and the trilateration method used. 

\begin{table}[t]
\footnotesize
\centering
\begin{tabular}{c c|c|c|c|c|c|} 
\cline{2-7}
& \multicolumn{1}{|c|}{8:55-9:10} & 9:10-9:25 & 9:25-9:40 & 9:40-9:55 & 9.55-10:10 & Avg. error \\ 
\hline
\multicolumn{1}{|c|}{Raw values} & 154.12  &  280.32  &  191.03  & 211.87 & 311.31 & 229.73 \\ 
\hline
\multicolumn{1}{|c|}{Look-back-5} & 154.13  &   283.61  &  190.65 & 212.37 & 305.06 & 229.16 \\ 
\hline
\multicolumn{1}{|c|}{Look-back-10} & 154.14  &  287.62   & 190.18  & 213.06 & 299.81 & 228.96 \\ 
\hline
\multicolumn{1}{|c|}{Look-back-15} & 154.15 &  291.62   &  189.71  & 213.73 & 294.82 & 228.81 \\
\hline
\multicolumn{1}{|c|}{Look-back-20} & 154.11  & 295.64   & 189.25   & 214.39 & 290.17 & 228.71 \\
\hline
\multicolumn{1}{|c|}{Look-back-30} & 153.87  & 304.67   & 188.39  & 215.52 & 274.92 & 227.47 \\
\hline
\multicolumn{1}{|c|}{Look-back-50} & 153.17  &  325.41   & 186.89  & 217.15 & 231.23 & 222.77 \\
\hline
\multicolumn{1}{|c|}{Kalman filter} & 154.4  & 405.09  & 174.83 & 235.13 & 141.67 & 222.22 \\ 
\hline
\multicolumn{1}{|c|}{Kalman filter +} & \multirow{2}{*}{154.4}  &  \multirow{2}{*}{405.56}  &  \multirow{2}{*}{174.74}  & \multirow{2}{*}{235.36} &  \multirow{2}{*}{143.63}  &  \multirow{2}{*}{222.74} \\ 
\multicolumn{1}{|c|}{look-back-5} &   &    &    & & &  \\ 
\hline
\multicolumn{1}{|c|}{Kalman filter +} & \multirow{2}{*}{154.4}  &  \multirow{2}{*}{406.14}  &  \multirow{2}{*}{174.62}  & \multirow{2}{*}{235.62} &  \multirow{2}{*}{145.99}  &  \multirow{2}{*}{223.35} \\ 
\multicolumn{1}{|c|}{look-back-10} &   &    &    &  & & \\ 
\hline
\multicolumn{1}{|c|}{Kalman filter +} & \multirow{2}{*}{154.4}  &  \multirow{2}{*}{406.69}  &  \multirow{2}{*}{174.51}  & \multirow{2}{*}{235.88} &  \multirow{2}{*}{148.27}  &  \multirow{2}{*}{223.95} \\ 
\multicolumn{1}{|c|}{look-back-15} &   &    &    &  & & \\ 
\hline
\multicolumn{1}{|c|}{Kalman filter +} & \multirow{2}{*}{154.4}  &  \multirow{2}{*}{407.23}  &  \multirow{2}{*}{174.40}  & \multirow{2}{*}{236.12} &  \multirow{2}{*}{150.46}  &  \multirow{2}{*}{224.52} \\ 
\multicolumn{1}{|c|}{look-back-20} &   &    &    &  & & \\ 
\hline
\multicolumn{1}{|c|}{Kalman filter +} & \multirow{2}{*}{154.37}  &  \multirow{2}{*}{408.24}  &  \multirow{2}{*}{174.2}  & \multirow{2}{*}{236.57} &  \multirow{2}{*}{154.56}  &  \multirow{2}{*}{225.59} \\ 
\multicolumn{1}{|c|}{look-back-30} &   &    &    &  & & \\ 
\hline
\multicolumn{1}{|c|}{Kalman filter +} & \multirow{2}{*}{154.33}  &  \multirow{2}{*}{410.02}  &  \multirow{2}{*}{173.87}  & \multirow{2}{*}{237.33} &  \multirow{2}{*}{161.57}  &  \multirow{2}{*}{227.42} \\ 
\multicolumn{1}{|c|}{look-back-50} &   &    &    &  & & \\ 
\hline
\end{tabular}
\caption{Localization errors for office location. Errors computed as averaged Euclidean distances, reported in centimetres.}
\label{table:second_experiment}
\end{table}

The results obtained using the neural network lead to conclusions similar to those obtained within the first location: generally, more layers and more training epochs mean improved accuracy, as illustrated in Table \ref{table:second_experiment_ann}. The largest error, obtained by a network with just one hidden layer and after only 100 training epochs is less than half the average error obtained by methods based on trilateration, while the smallest one is less than 8 centimetres. The average error for the conducted experiments is $35.23$cm and the $95\%$ confidence interval is $35.23 \pm 11.86$cm ($[23.36, 47.10]$). The localization accuracy obtained by the network is good and training times were lower than 1 minute on a computer with an Intel i5 CPU and 8 GB RAM, with all computations running on the CPU. However, the network is trained for 5 exact locations using approximately 2000 records. The main disadvantage of this method, as remarked for the first experiment, is that training it for a larger set of locations in a dwelling would require significantly more time, both for the measurements and for the training process. The results obtained by the best performing network are presented in Figure \ref{fig:loc_dwelling_office}. As seen in the figure, the predicted positions are very close to the real ones and many of them overlap.

\begin{table}[t]
\footnotesize
\centering
\begin{tabular}{c c|c|c|c|c|}
\cline{2-6}
& \multicolumn{5}{|c|}{Number of hidden layers} \\
\hline
\multicolumn{1}{|c|}{Epochs} & 1 & 2 & 3 & 4 & 5 \\ 
\hline
\multicolumn{1}{|c|}{100} & $106.98 \pm 15.11$ & $96.73 \pm 23.50$  & $77.57 \pm 22.40$ & $70.79 \pm 32.97$ & $51.75 \pm 9.77$ \\ 
\hline
\multicolumn{1}{|c|}{500} & $84.36 \pm 23.08$ & $46.05 \pm 18.69$  & $26.42 \pm 12.30$ & $19.16 \pm 4.55$ & $15.14 \pm 3.71$ \\ 
\hline
\multicolumn{1}{|c|}{1000} & $54.45 \pm 17.40$ & $30.23 \pm 14.54$  & $15.48 \pm 8.04$ & $11.16 \pm 4.03$ & $10.68 \pm 5.18$ \\ 
\hline
\multicolumn{1}{|c|}{2000} & $38.67 \pm 13.38$ & $23.44 \pm 13.13$ & $10.82 \pm 3.56$ & $10.89 \pm 3.99$ & $8.48 \pm 3.18$ \\ 
\hline
\multicolumn{1}{|c|}{3000} & $27.44 \pm 13.13$ & $16.54 \pm 11.47$  & $10.20 \pm 5.55$  &  $9.68 \pm 7.40$ & $7.73 \pm 4.08$ \\ 
\hline
\end{tabular}
\caption{Office location. Average error based on Euclidean distance $\pm$ standard deviation obtained using the artificial neural network, over 10 runs, with stratified cross-validation. Errors reported in centimetres.}
\label{table:second_experiment_ann}
\end{table}

\subsection{Comparison with Related Work}
Authors of \cite{bahl2000} present RADAR, a system that uses the strength of a 2.4Ghz WiFi signal from the \textit{k}-nearest neighbour devices to detect location using RSSI. The system was tested in a 980$m^2$ floor space of a multi-storey building divided into around 50 rooms. Experimental evaluation showed the system to have a resolution of 3 meters, enough for room-level detection. 

A similar result was achieved by Battiti et al. \cite{battiti2002}, who employ WiFi signal and a neural network for person positioning, and achieve similar, room-level resolution. Fariz et al. employ trilateration together with Kalman filters for determining the person's location \cite{fariz2018}. The error is between 3.58 and 14.82 meters. Deng et al. applied the extended Kalman filter based on fusing WiFi signals \cite{deng2015}, lowering the localization error to 2.83 meters. Another experiment regarding indoor positioning \cite{li2018} obtained an error equal to 3.29 meters when trilateration was used with Bluetooth and 1.61 meters when using a backpropagation neural network. 

In Sadowski et al., RSSI-based indoor localization data are collected using WiFi for two scenarios and the reported positioning errors were low \cite{sadowski2018}. The first scenario consisted of one large room in which a few BLE gadgets and different WiFi systems were placed in order to introduce signal interference. The second scenario involved a smaller room with furniture and no gadgets. In this case, a low degree of commotion was present. The average positioning error for the first scenario was 0.84 meters, and for the second scenario it reached 0.48 meters. The nodes used for RSSI value estimation were placed on several tables with similar height, restricting the quantity of signal reflections and interference.

In the case of long narrow spaces, Gao et al. \cite{gao2019} used same line dual connection for estimating the person location, because trilateration caused tracking difficulties. This solution considers the loss and gain of RSSI values both for transmitter and receiver. The reported accuracy of this solution was of 1.6 meters.

When compared with most relevant existing systems, our solution integrates additional capabilities into a cost-effective device that can be easily deployed and which looks unobtrusive in a typical home or office setting. In our experiments, the luminaires were ceiling-mounted and indoor localization was attempted in five rooms using trilateration combined with filter-based approaches. The complexity of carrying out a controlled evaluation of these technologies means that a detailed comparison with similar approaches cannot be made. Differences in enclosure sizes, room layouts, building materials, presence of large furniture or people, as well as signal interference from nearby devices cannot fully be accounted for outside a laboratory setting.

\section{Conclusion}
The trend of population ageing started several decades ago and recent data shows that over medium and long term, it is expected to amplify. The more concerning aspect however is that this process is not accompanied by an improved quality of life for older adults \cite{who15}. Our research evaluates indoor localization accuracy, an important aspect for an unobtrusive and cost-effective system designed to help older adults live within their own homes. To achieve our goals, the system hardware was implemented in the form of intelligent luminaires that replace existing lighting infrastructure. In addition to environment sensing \cite{marin2018}, they provide indoor localization using RSSI from a Bluetooth Low Energy device, such as a standard smartphone or wearable. 

We extended our initial evaluation \cite{marin19} to include two different, but representative locations for an indoor positioning system. We employed direct trilateration-based methods with the minimum number of required beacons and evaluated the accuracy of several post-processing approaches, including Kalman filtering as well as our proposed look-back-k heuristic. We combined the two approaches to verify if an initial noise reduction step improves our heuristic's accuracy. Finally, we trained an artificial neural network to verify whether the expected future location of the monitored person can be accurately determined. While not an actual localization problem, we believe it to represent an important step towards behavioural pattern recognition and profiling, which existing research \cite{soto2015,kim2010} suggests can be important for early detection of several medical conditions related to ageing.

Our evaluation shows that direct trilateration is suitable for achieving room-level localization accuracy in realistic scenarios. However, raw readings presented significant variation. When studying the obtained distance estimations, we observe that in several situations a common intersection could not be identified, resulting in large error margins. This situation arises both because of noisy RSSI readings caused by signal interference, as well as multipath fading and environmental factors. We plan to also experiment with other, more complex algorithms, for instance with non-linear least squares optimisation, to compensate for noisy data. In the undertaken experiments, all post-processing approaches improved the observed accuracy, starting with the Kalman noise reduction filter. In both evaluated scenarios, improved results were obtained when the Kalman filter was combined with our look-back heuristic. 

When compared to other indoor positioning systems, our proposed approach does not require the installation of additional devices or wiring, and as shown, can achieve room-level accuracy in realistic scenarios that include signal pollution. Its deployment is straightforward and involves replacing some of the existing light bulbs with intelligent luminaires. A comparative evaluation of localization accuracy is however not feasible, as it would require a controlled but realistic environment where signal pollution can be controlled and where multiple systems could be deployed. As such, we limited ourselves to comparing each system's prerequisites and the obtained results.

With regards to future work, in addition to evaluating more complex algorithms for indoor localization, we aim to carry out an evaluation on the accuracy that can be achieved when tracking multiple targets, as well as moving subjects. In both cases, luminaires will have to increase signal strength polling rate. Accurate indoor localization of moving subjects will better help detect older adult behaviour patterns and identify risky situations, as well as enabling other functionalities in both private and public places.

\subsubsection*{\uppercase{Acknowledgement}}
\noindent This work was supported by a grant of the Romanian National Authority for Scientific Research and Innovation, CCCDI UEFISCDI, project number 46E/2015, \emph{i-Light - A pervasive home monitoring system based on intelligent luminaires}.

\bibliographystyle{splncs03}
\bibliography{bibliography}

\end{document}